# Nonlinear dispersion relation in anharmonic periodic mass-spring and mass-in-mass systems

R. Zivieri[1,2,#], F. Garescì[3,#,*], B. Azzerboni,[3] M. Chiappini[4], G. Finocchio[1,4]

[1]Department of Mathematical and Computer Sciences, Physical Sciences and Earth Sciences, University of Messina, Italy
[2]Istituto Nazionale di Alta Matematica (INdAM), Roma, Italy
[3]Department of Engineering, University of Messina, Italy
[4]Istituto Nazionale di Geofisica e Vulcanologia (INGV), Roma, Italy

**Abstract**

The study of wave propagation in chains of anharmonic periodic systems is of fundamental importance to understand the response of dynamical absorbers of vibrations and acoustic metamaterials working in nonlinear regime. Here, we derive an analytical nonlinear dispersion relation for periodic chains of anharmonic mass-spring and mass-in-mass systems resulting from considering the hypothesis of weak anharmonic energy and a periodic distribution function as ansatz of a general solution of the nonlinear equations of motion. Numerical simulations show that this expression is valid for anharmonic potential energy up to 50% of the harmonic one. This work provides a simple tool to design and study nonlinear dynamics for a class of seismic metamaterials.

# 1. Introduction

The study of systems including anharmonic springs has a fundamental importance in several applications of nonlinear mechanics such as vibration controls[1] [2], acoustic metamaterials[3] and phononics. The first attempt to study a nonlinear mass-spring system dates back to Fermi-Pasta-Ulam.[4] In particular, the first works studied one-dimensional dynamical systems of particles interacting with cubic forces between first-neighbors arising from an anharmonic potential[4],[5] and reproduced nonlinear phonon behavior in real crystal lattices observed by means of neutron scattering technique.[6] Anharmonic interactions of many-body potentials are also crucial to quantitatively reproduce cohesive and sublimation energy and lattice parameter behavior as a function of temperature in crystal lattices.[7],[8] Recently the study of nonlinear periodic systems has been faced in several works,[9] for instance considering single nonlinear partial differential equation via a continuum approximation,[10] variational equations,[11] and perturbative approaches.[12] On the other hand, after the experimental demonstration of locally resonant sonic metamaterials [13] there has been a growing interest to apply those concepts in the field of seismic metamaterials in order to filter seismic waves [14],[15],[16]. Most of the recent studies on this class of metamaterials deal with linear dynamics and some approaches are based on the use of one-dimensional models such as mass-in-mass systems or phononic crystals [16],[17],[18],[19]. Seismic metamaterials can be designed to be integrated directly within the soil,[14],[20] or, with the standard foundation of a building to protect via the concept of the composite foundation. The latter has the advantage that the physical space required for the isolation is comparable with the size of the building itself and from a modeling point of view can be described by a periodic chain of mass-in-mass systems [16]. In addition, compression tests show that the curve stress-strain of the rubber used for the external spring exhibits a nonlinear behavior that can be characterized by an anharmonic energy term. The main motivation of this work is to understand how the dynamical properties of composite foundation depends on this nonlinearity.

To do that, we derived an analytical nonlinear dispersion relation in periodic chain of mass-spring

(Fig. 1a) and mass-in-mass (Fig. 1b) systems in the presence of anharmonic potential [21] that in general can have broader applications. We have benchmarked the analytical expression with numerical simulations finding that it can be used up to wave amplitude giving anharmonic energy near half of harmonic one. We also show how the analytical expression can be generalized to systems including anharmonic potential expanded to all even orders (reciprocal behavior of wave propagation). This methodology differs from the perturbation approach [22] and the harmonic balance approach [23] already proposed in other works. In addition, our theoretical approach is different from the study reported by Fang *et al* [24] where similar non-linear dynamical systems, having the anharmonic energy for the internal spring, have been investigated via the harmonic average approach, continuation algorithm, Newton method and Lyapunov exponents.

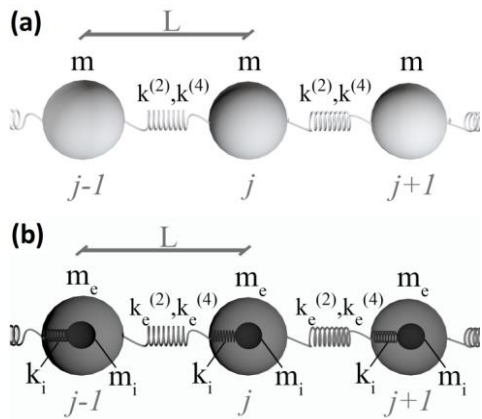

**Fig. 1**. A sketch of the periodic chain of (a) mass-spring and (b) mass-in-mass systems studied in this work. The letter *j* refers to the *j*-element while *L* is the spatial periodicity.

## 2. Experimental characterization of the rubber

In this Section, we present the stress-strain analysis of a nonlinear rubber that is performed by using a quasi-static compression test with deformation velocity equal to 80 mm/s and a maximum displacement equal to 45% of the thickness of the sample. Fig. 2 shows the restoring force as a function of the displacement amplitude for one sample (similar data have been obtained for other five different samples) in magenta circles. The equivalent constant stiffness $k_e$ increases as a function of the displacement exhibiting a significant nonlinear trend for amplitudes larger than 0.01m. The result of the numerical interpolation of the experimental data according to a third-order polynomial equation, $F = k_e^{(2)}x + k_e^{(4)}x^3$, where $x$ is the displacement, is achieved with $k_e^{(2)} = 114 \text{ kN/m}$ and $k_e^{(4)} = 200 \text{ MN/m}^3$ (the order of magnitude of the fitting parameters is the same as the other samples) is also reported in Fig. 2 with a solid line. Those data have been used for the numerical test of the nonlinear analytical dispersion relation derived ahead in the text.

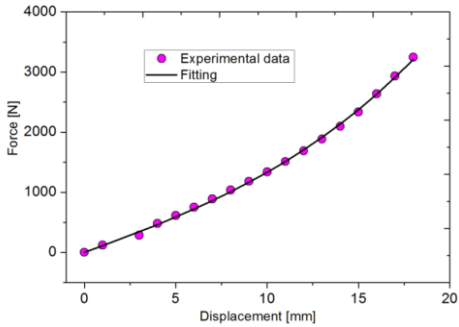

**Fig. 2.** Experimental data of a compression test on a sample of the rubber used for the realization of the composite foundation in [15], showing the restoring force as a function of the displacement amplitude. Magenta circles are the experimental data while the solid line is the fitting with a third-order polynomial equation $F = k_e^{(2)}x + k_e^{(4)}x^3$ (the fitting parameters are $k_e^{(2)} = 114 \text{ kN/m}$ and $k_e^{(4)} = 200 \text{ MN/m}^3$).

### 3. Theoretical approach and analytical development

This Section outlines the fundamental steps for the derivation of the nonlinear analytical dispersion relation. It is well known, that the linear dispersion relationship for the periodic chain of mass-spring (p.m.s.) and mass-in mass (p.m.m.) are given by the solution of the following equations:

$$m\omega^2 - 2k^{(2)}(1 - \cos(\kappa L)) = 0, \tag{1a}$$

and:

$$m_e m_i \omega^4 - \left(k_i(m_e + m_i) + 2m_i k_e^{(2)}(1-\cos(\kappa L))\right)\omega^2 + 2k_e^{(2)} k_i (1-\cos(\kappa L)) = 0, \quad (1b)$$

where $\omega$ is the angular frequency and $\kappa$ is the wave vector. $m$ and $k^{(2)}$ are the mass and the elastic constant of the mass-spring system, while $m_e$ ($m_i$) and $k_e^{(2)}$ ($k_i$) are the mass and the elastic constant of the mass-in-mass system, the subscripts $i$ and $e$ refer to the internal and external element, respectively (see Fig. 1 of the main text). The upper script $^{(2)}$ is used to indicate the harmonic elastic constant. In the linear response, the dispersion relations do not depend on the wave amplitude. In particular, the p.m.s. is characterized by an acoustic branch and a Bragg region for $\omega > \sqrt{\dfrac{2k^{(2)}}{m}}$, while the dispersion relation for the p.m.m. has an acoustic and an optical branch separated by a bandgap and a Bragg region for frequencies larger than the maximum frequency of the optical branch [17],[18].

Let start with the derivation of a nonlinear dispersion relation for the p.m.s. system where the spring is modeled by an anharmonic potential. The Lagrangian is given by

$$\mathcal{L} = \frac{1}{2} m (\dot{s}^j)^2 - \left[ \frac{1}{2} k^{(2)} (s^j - s^{j-1})^2 + \frac{1}{2} k^{(2)} (s^{j+1} - s^j)^2 + \frac{1}{4!} k^{(4)} \left[ (s^j - s^{j-1})^4 + (s^{j+1} - s^j)^4 \right] \right], \quad (2)$$

where $k^{(4)}$ is the 4$^{\text{th}}$-order anharmonic spring constant, while the symbols $s^j$, $s^{j+1}$ and $s^{j+1}$ are the mass displacements at the $j^{\text{th}}$, $(j+1)^{\text{th}}$ and $(j-1)^{\text{th}}$ element, respectively. Considering the hypothesis of weak anharmonic energy, [25],[26] i.e. the power of the higher harmonics terms is smaller than the fundamental frequency, the solution at different elements can be linked with a good approximation as $s^{j \pm 1}(t) = u^j(t) \exp(\pm i \kappa L)$ being $i$ the imaginary unit and $\kappa$ the spatial frequency (wave vector),[5] the dynamical equation can then be written as:

$$m \ddot{u}^j + 2k^{(2)}(1-\cos(\kappa L))u^j + \frac{1}{3!} k^{(4)} [2 - 2\cos(3\kappa L) + 6\cos(2\kappa L) - 6\cos(\kappa L)](u^j)^3 = 0. \quad (3)$$

The Fourier Transform of Eq. (3) reads:

$$-m\omega^2 U^j + 2k^{(2)}(1-\cos(\kappa L))U^j + \frac{1}{3!}k^{(4)}[2-2\cos(3\kappa L)+6\cos(2\kappa L)-6\cos(\kappa L)]\,\mathcal{F}\left[(u^j)^3\right]=0. \quad (4)$$

being $\mathcal{F}(\cdot)$ the Fourier operator and $U^j$ is the Fourier transform of $u^j$. Now we consider a family of non-trivial solutions $u^j$ that are distributions[27] having the property that, when raised to any exponent $n \in \mathbb{N}$, $(u^j)^n$ is proportional to $u^j$. In particular, we have considered, as ansatz, periodic rectangular bipolar pulses:

$$u^j = \mathrm{rect}(nt/T)_{n\in\mathbb{N}} = \begin{cases} -A_0 & \text{if } (n-1)T \le t < (n-1)T + \frac{T}{2} \\ A_0 & \text{if } (n-1)T + \frac{T}{2} \le t < nT \end{cases}, \quad (5)$$

where $A_0$ and $T$ are the amplitude and the period respectively. By substituting Eq. (5) in Eq. (4), we derive the following nonlinear dispersion relation that links the frequency ($\omega/2\pi = 1/T$), wave vector $\kappa L$ and wave amplitude $A_0$:

$$m\omega^2 - 2k^{(2)}(1-\cos(\kappa L)) - 2A_0^2\, k^{(4)} f_4(\kappa L) = 0, \quad (6)$$

where $f_4(\kappa L) = \frac{1}{3!}[1-\cos(3\kappa L)+6\cos(2\kappa L)-6\cos(\kappa L)]$. The same approach can be applied to the p.m.m. system. The dynamical equations are:

$$\begin{cases} m_e \ddot{u}_e^j + k_e^{(2)}(2-2\cos(\kappa L))u_e^j + k_i(u_e^j - u_i^j) + \frac{1}{3!}k_e^{(4)}[2-2\cos(3\kappa L)+6\cos(2\kappa L)-6\cos(\kappa L)](u_e^j)^3 = 0 \\ m_i \ddot{u}_i^j + k_i(u_i^j - u_e^j) = 0 \end{cases}, \quad (7)$$

while the nonlinear dispersion relation is given by the following equation:

$$m_e m_i \omega^4 - \left(k_i(m_e+m_i)+2m_i k_e^{(2)}(1-\cos(\kappa L))+m_i A_0^2 f_4(\kappa L)\right)\omega^2 \\ +2k_e^{(4)} k_i(1-\cos(\kappa L)+k_i A_0^2 f_4(\kappa L)) = 0. \quad (8)$$

Eq. (6) and Eq.(8) are the main result of this manuscript. They represent an extension of the linear dispersion relation in the presence of an anharmonic potential. Because we are not in the harmonic regime, the frequency, the wave vector and the wave amplitude of the nonlinear dispersion relation are related to the first harmonics or fundamental frequency of the wave.

It can be easily demonstrated that the results are robust for a wide range of trial solutions. More generally, one calculates $u^j$ and $(u^j)^3$ or $u_e^j$ and $(u_e^j)^3$ considering distributions satisfying the property of the ansatz in Eq. (5) and, before applying the Fourier Transform, both terms are expanded in term of their Fourier series. Finally, it can be derived Eq. (6) and Eq.(8) for each term of the Fourier series. With this procedure, it is trivial to find for the distribution family of Eq. (5) (defined for any $T$ and any $A_0$) a valid Eq. (6) and Eq.(8) for all $sin(2n\pi t/T)$ $n \in \mathbb{N}$ functions, that now should be all verified. The same argument can be extended to the $cos(2n\pi t/T)$ functions if it is introduced a proper time shifting for the rectangular bipolar pulses.

## 4. Numerical Solution of the dynamical equations and benchmark of the nonlinear analytical dispersion relation

This Section is mainly devoted to the numerical numerical test of nonlinear dispersion relation derived for the p.m.s. (p.m.m.) systems. In order to study the validity of the ansatz solution expressed in Eq.(5), we numerically solve the differential equations for the p.m.s. (p.m.m.) systems with Heun time integration scheme [28] considering as boundary conditions a sinusoidal input displacement at the first cell $s^1 = A_0 \sin(\omega t)$ ($s_e^1 = A_0 \sin(\omega t)$) and a chain long enough in order to avoid any reflection of the wave (all the numerical results presented here are for simulations long 30s and considering $N=400$ elements of the periodic chains) [29]. The numerical solutions of the dynamical equations for the periodic chain of mass-spring systems [29] is obtained by using the procedures described below.

*4.1. Numerical model for the mass-spring system*

The nonlinear differential equations describing the time domain evolution of the displacement at each cell for the periodic chain of mass-spring system that includes also viscous dissipative element for each cell is given by:

$$m\frac{d^2 s^j}{dt^2} + b\left(2\frac{ds^j}{dt} - \frac{ds^{j+1}}{dt} - \frac{ds^{j-1}}{dt}\right) + k^{(2)}\left(2s^j - s^{j+1} - s^{j-1}\right) + \frac{1}{3!}k^{(4)}\left[(s^j - s^{j-1})^3 + (s^j - s^{j+1})^3\right] = 0, \quad (9)$$

where the coefficient $b$ is the mass and the viscous damping coefficient of the system (the same for all the chain). Eq. (9) can be rewritten in terms of a finite difference scheme. In our implementation we consider $N$ elements, and the anharmonic term is taken into account only from the 2nd to the $(N-1)$th elements:

$$\begin{cases} m\dfrac{d^2 s^1}{dt^2}+b\left(\dfrac{ds^1}{dt}-\dfrac{ds^2}{dt}\right)+k^{(2)}\left(s^1-s^{j+1}\right)=0 \\ m\dfrac{d^2 s^j}{dt^2}+b\left(2\dfrac{ds^j}{dt}-\dfrac{ds^{j+1}}{dt}-\dfrac{ds^{j-1}}{dt}\right)+k^{(2)}\left(2s^j-s^{j+1}-s^{j-1}\right)+\dfrac{1}{3!}k^{(4)}\left[(s^j-s^{j-1})^3+(s^j-s^{j+1})^3\right]=0. \\ m\dfrac{d^2 s^N}{dt^2}+b\left(\dfrac{ds^N}{dt}-\dfrac{ds^{N-1}}{dt}\right)+k^{(2)}\left(s^N-s^{N-1}\right)=0 \end{cases} \quad (10)$$

For each $j$, Eq. (10) is written as a system of differential equations of the first order:

$$\begin{cases} \dfrac{ds^j}{dt}=v_e^j \\ \dfrac{d^2 s^j}{dt^2}=\dfrac{dv_e^j}{dt} \end{cases}. \quad (11)$$

The wave is excited in the chain considering a controlled displacement at the first element having a fixed frequency $s^1 = A\sin(\omega t)$. All the initial conditions are set to zero. The damping term $b$ has been introduced to have better stability of the numerical algorithm and it has been set equal to 0.2.

*4.2. Numerical model for the mass-in-mass system*

The nonlinear differential equations describing the time domain evolution of the displacement of the internal and external mass for the mass-in-mass system are given by:

$$\begin{cases} m_e \dfrac{d^2 s_e^j}{dt^2}+b_e\left(2\dfrac{ds_e^j}{dt}-\dfrac{ds_e^{j+1}}{dt}-\dfrac{ds_e^{j-1}}{dt}\right)+k_e^{(2)}\left(2s_e^j-s_e^{j+1}-s_e^{j-1}\right)+k_i\left(s_e^j-s_i^j\right)+b_i\left(\dfrac{ds_e^j}{dt}-\dfrac{ds_i^j}{dt}\right) \\ \qquad +\dfrac{1}{3!}k^{(4)}\left[(s_e^j-s_e^{j-1})^3+(s_e^j-s_e^{j+1})^3\right]=0 \\ m_i \dfrac{d^2 s_i^j}{dt^2}+k_i\left(s_i^j-s_e^j\right)+b_i\left(\dfrac{ds_i^j}{dt}-\dfrac{ds_e^j}{dt}\right)=0 \end{cases} \quad (12)$$

where the coefficients $b_e$, and $b_i$ are the external and internal viscous damping coefficients of the system. Eq. (12) can be rewritten in term of a finite difference scheme similarly to Eq. (10) (not shown here). We consider the same boundary conditions as in the case of periodic chain of mass-spring system (Section 4.1). The damping terms $b_e$ and $b_i$ are set equal to 0.1 and 0.5 for a better numerical stability.

## 5. Results

The results of a comparison between the analytical model developed in this work (Eq. (6)) and the numerical calculations for the p.m.s. system achieved for $m = 245$ Kg, $k^{(2)} = 155$ kN/m and $k^{(4)} = 200$ MN/m³ and for different wave amplitude ($A_0 = 1$ mm, (b) $A_0 = 2$ mm, (c) $A_0 = 5$ mm, (d) $A_0 = 7$ mm, (e) $A_0 = 10$ mm, and (f) $A_0 = 20$ mm) are shown in Fig. 3. The parameters are typical experimental values for spring with low Young Modulus[16]. For this set of parameters, we found a quantitative agreement up to 10 mm (Fig. 3(e)) and a semi-quantitative agreement up to $A_0 = 20$ mm (Fig. 3(f)). A systematic comparison in a wide set of parameters drives us to the empirical conclusion that, the hypothesis of weak anharmonic energy used to derive Eq. (6) is valid for wave amplitude $A_0^2 < 0.5 \frac{k^{(2)}}{k^{(4)}}$ ($\sqrt{\frac{k^{(2)}}{k^{(4)}}} \approx 28$ mm), however we can still use Eq. (6) to have a nonlinear dispersion for a semi-quantitative description of the wave propagation up to $A_0 < 0.7 \sqrt{\frac{k^{(2)}}{k^{(4)}}}$. Similar results are obtained for the case of soft spring (i.e. $k^{(4)} < 0$). In order to test the robustness of the model, we have also performed a comparison considering also soft spring, i.e. $k^{(4)} < 0$. Fig. 4 summarizes the results of the comparison for $m = 245$ Kg, $k^{(2)} = 155$ kN/m and $k^{(4)} = -200$ MN/m³.

**Commentato [g1]:** Verificare se corretto.

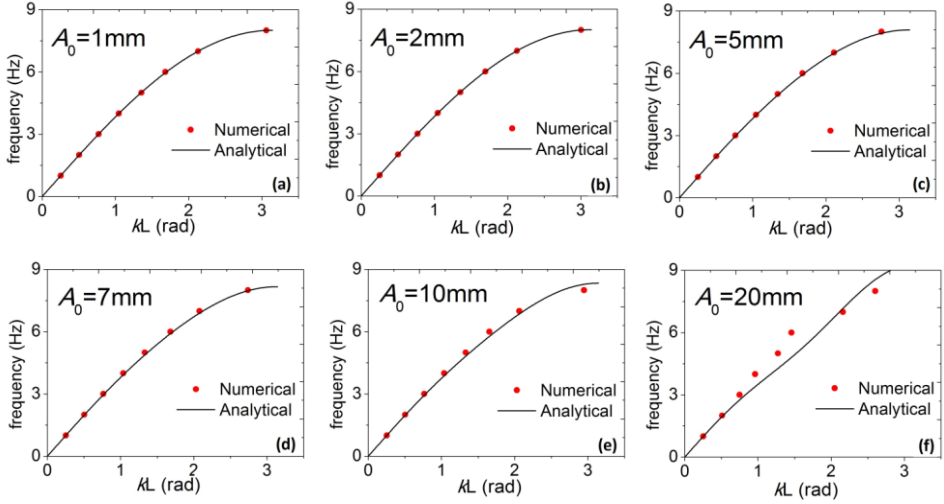

**Fig. 3**. A comparison of the analytical nonlinear dispersion relation of Eq. (6) (solid line) with the numerical calculations (circles) for different wave amplitudes: (a) $A_0=$ 1 mm, (b) $A_0=$ 2 mm, (c) $A_0=$ 5 mm, (d) $A_0=$ 7 mm, (e) $A_0=$ 10 mm, and (f) $A_0=$ 20 mm.

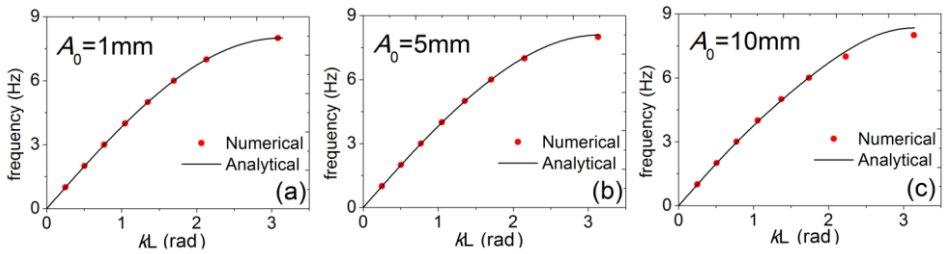

**Fig. 4**. A comparison of the analytical nonlinear dispersion relation of Eq. (6) (solid line) of the main text with the numerical calculations (circles) for different wave amplitudes: (a) $A_0=$ 1 mm, (b) $A_0=$ 5 mm, and (c) $A_0=$ 10 mm.

To summarize the results, for a fixed frequency the main effect due to anharmonicity is the increasing of the wave vector $\kappa L$ as a function of the wave amplitude and this shift can be computed analytically from Eq. (6).

We also performed the same comparison for the p.m.m. systems, numerical computations and Eq. (8), that is summarized in Fig. 5 ($m_e$ = 245 Kg, $k_e^{(2)}$ = 155 kN/m and $k_e^{(4)}$ = 200 MN/m³, $k_i$ =1080 kN/m, $m_i$ = 317 Kg) for different wave amplitudes: (a) $A_0$= 1 mm, (b) $A_0$= 5 mm, and (c) $A_0$= 10 mm. Also in this case, the effect due to anharmonicity is the shift of the wave vector $\kappa L$ as a function of the wave amplitude at fixed frequency especially for the low-frequency acoustic branch.

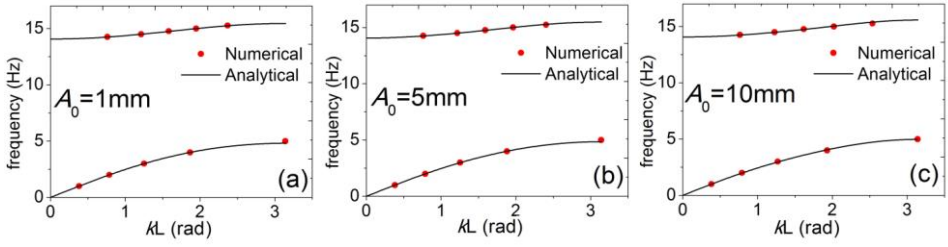

**Fig. 5**. A comparison of the analytical nonlinear dispersion, both acoustic and optical branches, relation of Eq. (8) (solid line) with the numerical calculations (circles) for different wave amplitudes: (a) $A_0$= 1 mm, (b) $A_0$= 5 mm, and (c) $A_0$= 10 mm.

## 6. Dispersion relation in the presence of an anharmonic potential of any even order

We have also developed the formalism to generalize the dispersion relation in the presence of an anharmonic potential of any even order (reciprocal and chiral systems). The general expression for the anharmonic potential (reciprocal and passive systems) $V^j$ at the $j$th-element is given by:

$$V^j = \left(V_0 + \frac{1}{2}k^{(2)}(s^j - s^{j-1})^2 + \frac{1}{2}k^{(2)}(s^{j+1} - s^j)^2\right) + \sum_{n=2}^{\infty} k^{(2n)} \frac{1}{2n!}\left[(s^j - s^{j-1})^{2n} + (s^{j+1} - s^j)^{2n}\right], \quad (13)$$

where $k^{(2n)}$ is the $2n^{th}$ anharmonic spring constant. The dynamical equation for the $j^{th}$-element can be computed via the anharmonic Lagrangian $\mathcal{L} = \frac{1}{2}m\left(\dot{s}^j\right)^2 - V^j$ as $\frac{d}{dt}\left(\frac{\partial \mathcal{L}}{\partial \dot{s}^j}\right) - \frac{\partial \mathcal{L}}{\partial s^j} = 0$, and it is given by the following equation:

$$m\ddot{s}^j - Q^j = 0, \quad (14)$$

where $Q^j = -\sum_{n=1}^{\infty} k^{(2n)} \frac{1}{(2n-1)!}\left[\left(s^j - s^{j-1}\right)^{2n-1} - \left(s^{j+1} - s^j\right)^{2n-1}\right]$. Eq. (14) is derived without any approximation and, in general, the solution $s^j$ should have a complex time and spatial dependence $s^j(t, s^{j+1}, s^{j-1})$.

*6.1. Working hypotheses*

The solution has been obtained according to two main hypotheses, the spatial factorization of the solution and the use of a proper ansatz for the solution of Eq. (14). For the first hypothesis, i.e. the factorization of the solution at the $j$-element, we have postulated that there exists a spatial translation function $f(\pm \kappa L)$, having a generic dependence on $\kappa L$, that links the solution in adjacent cells, $s^{j\pm 1}(t) = u^j(t) f(\pm \kappa L)$, where $u^j(t)$ is the function describing the time dependent part of the solution and at the $j^{\text{th}}$ element coincides with $s^j(t)$. Under this hypothesis, Eq. (14) reads:

$$m \ddot{u}^j + \sum_{n=1}^{\infty} k^{(2n)} \, f^{(2n)} \, \left(u^j\right)^{2n-1} = 0, \qquad (15)$$

where $f^{(2n)}(\kappa L) = \frac{1}{(2n-1)!}\left[\left(1 - f(-\kappa L)\right)^{2n-1} + \left(1 - f(\kappa L)\right)^{2n-1}\right]$. Now let us apply the Fourier Transform to Eq. (12) that leads:

$$-m\omega^2 U^j + \sum_{n=1}^{\infty} k^{(2n)} \, f^{(2n)} \, \mathcal{F}\left(\left(u^j\right)^{2n-1}\right) = 0. \qquad (16)$$

We now apply the second hypothesis considering the same ansatz of Eq. (5). Hence, by substituting Eq. (5) in Eq. (16), we derive the following expression:

$$\omega^2 U^j - \sum_{n=1}^{\infty} \frac{k^{(2n)} A_0^{2n-2}}{m} f^{(2n)} \, U^j = 0, \qquad (17)$$

that is valid for any $T$ and $A$, and hence the nonlinear dispersion relation takes the form:

$$\omega^2 = \sum_{n=1}^{\infty} \frac{k^{(2n)} A_0^{2n-2}}{m} f^{(2n)}, \qquad (18)$$

that represents a natural extension of the above linear dispersion relation. The expression for the anharmonic potential $V_j$ at the $j^{\text{th}}$-element is given by:

$$V^j = \left( V_0 + \frac{1}{2} k_i (s_i^j - s_e^j)^2 + \frac{1}{2} k_e^{(2)} (s_e^j - s_e^{j-1})^2 + \frac{1}{2} k_e^{(2)} (s_e^{j+1} - s_e^j)^2 \right) \\ + \sum_{n=2}^{\infty} \frac{1}{2n!} k_e^{(2n)} \left[ (s_e^j - s_e^{j-1})^{2n} + (s_e^{j+1} - s_e^j)^{2n} \right], \quad (19)$$

where $k^{(2n)}$ is the $2n^{\text{th}}$ anharmonic spring constant of the external spring, while $s_e^j$ and $s_i^j$ are the displacements of the external and internal mass, respectively. The dynamical equations can be written as follow:

$$\begin{cases} m_e \ddot{u}_e^j + k_i (u_e^j - u_i^j) + \sum_{n=1}^{\infty} k_e^{(2n)} \left( f^{(2n)} \right)^{2n-1} = 0 \\ m_i \ddot{u}_i^j + k_i (u_i^j - u_e^j) = 0 \end{cases}. \quad (20)$$

Applying the Fourier transform to Eq. (20), we get $U_i^j = \dfrac{k_i U_e^j}{k_i - m_i \omega^2}$ where $U_i^j$ and $U_e^j$ are the Fourier transform of $u_i^j$ and $u_e^j$, respectively and then the following equation:

$$\left[ m_e m_i \omega^4 - \left( k_i (m_e + m_i) + 2 m_i k_e (1 - \cos(\kappa L)) \right) \omega^2 + 2 k_e k_i (1 - \cos(\kappa L)) \right] U_e^j \\ + \left( k_i - m_i \omega^2 \right) \sum_{n=2}^{\infty} k_e^{(2n)} f^{(2n)} \mathcal{F}\left[ (u_e^j)^{2n-1} \right] = 0. \quad (21)$$

By considering the same ansatz of Eq. (5), the dispersion relation of the mass-in-mass system turns out to be:

$$\left[ m_e m_i \omega^4 - \left( k_i (m_e + m_i) + 2 m_i k_e^{(2)} (1 - \cos(\kappa L)) \right) \omega^2 + 2 k_e^{(2)} k_i (1 - \cos(\kappa L)) \right] \\ + \left( k_i - m_i \omega^2 \right) \sum_{n=2}^{\infty} k_e^{(2n)} A_0^{2n-2} f^{(2n)} = 0. \quad (22)$$

Within this formulation, one challenge will be to find out an expression for $f(\pm \kappa L)$ that is more general than the $\exp(\pm i \kappa L)$.

*6.2. Duffing equation for a periodic chain of mass-spring system*

It is well known that the cubic Duffing differential equation[30] is the standard equation governing mass-spring systems with fourth-order order anharmonic potential. Here, we show that the Duffing equation can describe the dynamics of the $j^{th}$ cell of a chain of mass-spring system under the first working hypothesis discussed in Section 6.1 according to which the general solution can be spatially factorized via the $f(\pm \kappa L)$ spatial factor in the presence of fourth-order anharmonic interactions, $s^{j\pm 1}(t) = u^j(t) f(\pm \kappa L)$.

Keeping in the sum on the second member of Eq.(15) the terms $n = 1, 2$ the equation of motion for the $j^{th}$ cell can be cast in the form of a non-linear and homogeneous third-order Duffing equation:

$$m \ddot{u}^j - \varepsilon \left(u^j\right)^3 + \eta \, u^j = 0, \qquad (23)$$

where $\varepsilon = \varepsilon(\kappa L)$ and $\eta = \eta(\kappa L)$. In particular, $\varepsilon = -k^{(4)} f^{(4)}$ is the nonlinear parameter and $\eta = k^{(2)} f^{(2)}$ is the linear parameter with $f^{(4)}(\kappa L) = 2 - \left((f(\kappa L))^3 + (f(-\kappa L))^3\right) - 3\left(f(-\kappa L) - (f(\kappa L))^2\right) + 3\left((f(-\kappa L))^2 - f(\kappa L)\right)$ and $f^{(2)}(\kappa L) = 2 - f(\kappa L) - f(-\kappa L)$.

Eq.(23) admits an exact analytical solution:

$$u^j(\kappa L, t) = A^j \, \mathrm{sn}(Bt \,|\, \bar{m}), \qquad (24)$$

where $\mathrm{sn}(\ )$ is the Jacobian elliptic sine with $A^j = \sqrt{\dfrac{2m c_1}{\eta + \sqrt{\eta^2 - 2m \varepsilon c_1}}}$ the anharmonic amplitude of the wave at the $j^{th}$ cell, $B = \sqrt{\dfrac{\eta + \sqrt{\eta^2 - 2m \varepsilon c_1}}{2m}}$ the argument (after setting, without loss of generality, the initial instant of time $t_0 = 0$) and $\bar{m} = \dfrac{\eta - \sqrt{\eta^2 - 2m \varepsilon c_1}}{\eta + \sqrt{\eta^2 - 2m \varepsilon c_1}}$ the parameter of the elliptic function with the integration constant $c_1 = 1$ m$^2$/s$^2$. Here, $A^j = A^j(\kappa L)$  $B = B(\kappa L)$ and $\bar{m} = \bar{m}(\kappa L)$. As $\varepsilon \to 0$, we get $u(\kappa,t)=\sqrt{\frac{mc_1}{\eta}}\sin(\omega_1 t)$, one of the solutions of the harmonic chain of mass-spring system corresponding to the first resonance frequency $\omega_1=\sqrt{\frac{\eta}{m}}$ of free vibration in the limit of small oscillations.

Eq.(23) and Eq.(24) are a generalization of the cubic Duffing equation, where the linear and cubic coefficients and its solution (amplitude, argument and parameter) contain information on spatial periodicity and on the propagation properties of the anharmonic wave.

## 7. Conclusion

We have introduced a method to determine a simple analytical expression for the nonlinear dispersion relation of the propagating wave in chains of anharmonic periodic mass-spring systems and mass-in-mass systems. The formula has been benchmarked with numerical simulations finding an agreement in the region of wave amplitude where the anharmonic strength is less than 50% of the harmonic one. These results can be used in nonlinear acoustic. In addition, due to the large energies released in the presence of earthquakes [31], the use of an anharmonic potential should be the first step to extend the study of wave propagation in the soil [15] in a weak nonlinear regime and to study the dynamical properties of some categories of seismic metamaterials [16] [32].

**Acknowledgments**

The present research has been supported by the Project (Progetto Premiale) "Strategic Initiatives for the Environment and Security -SIES". R.Z. acknowledges the support by National Group of Mathematical Physics (GNFM-INdAM) and Istituto Nazionale di Alta Matematica "F. Severi". The authors thank Domenico Romolo for the graphical support.